\documentclass[a4paper]{article}
\usepackage{graphicx}
\usepackage{onecolpceurws}
\usepackage{wrapfig}
\usepackage{amsmath}
\usepackage{blindtext}
\usepackage{subfig}
\usepackage{esvect}
\usepackage[utf8]{inputenc}
\usepackage{color}
\usepackage[square,sort,comma,numbers]{natbib}

\usepackage[numbers]{natbib}
\usepackage{amsfonts}
\usepackage{setspace}
\usepackage{lineno}
\usepackage{hyperref}
\usepackage{url}
\usepackage{commath}

\usepackage{indentfirst}
\title{Earthquake Detection in 1-D Time Series Data with \\Feature Selection and Dictionary Learning}

\author{Zheng Zhou, Youzuo Lin$^{*}$, Zhongping Zhang, Yue Wu, and Paul Johnson}
\institution{Earth and Environment Sciences Division, Los Alamos National Laboratory, Los Alamos, NM 87545}

\begin{document}
\maketitle

\begin{abstract}

Earthquakes can be detected by matching spatial patterns or phase properties  from 1-D seismic waves. Current earthquake detection methods, such as waveform correlation and template matching, have difficulty detecting anomalous earthquakes that are not similar to other earthquakes. In recent years, machine-learning techniques for earthquake detection have been emerging as a new active research direction. In this paper, we develop a novel earthquake detection method based on dictionary learning. Our detection method first generates rich features via signal processing and statistical methods, and further employs feature selection techniques to choose features that carry the most significant information. Based on these selected features, we build a dictionary for classifying earthquake events from non-earthquake events. To evaluate the performance of our dictionary-based detection methods, we test our method on a labquake dataset from Penn State University, which contains 3,357,566 time series data points with a 400 MHz sampling rate. 1,000 earthquake events are manually labeled in total, and the length of these earthquake events varies from 74 to 7151 data points. Through comparison to other detection methods, we show that our feature selection and dictionary learning incorporated earthquake detection method achieves a 80.1\% prediction accuracy and outperforms the baseline methods in earthquake detection, including Template Matching (TM) and Support Vector Machine (SVM).

\end{abstract}
\vskip 32pt

\section{Introduction}
\indent Earthquake detection, usually applied to 1-D time series seismic data, is one of the most important tasks in seismology. Detecting the earthquake events with different durations via the same detection algorithm still remains as a challenging task for the researchers who work on developing automated earthquake detection methods, because the features and characteristics of various earthquake events can vary significantly \citep{wu2018deepdetect,Gib-2006,Wit-1998}. The explosive development and successful application of machine learning techniques in many time series detection tasks have revealed that applying machine learning methods to earthquake detection tasks could be a promising direction. Among all these machine learning methods, dictionary learning has achieved striking results in signal denoising, signal compression, and classification based on time series data \citep{det-2015}.

Traditionally, while applying dictionary learning methods to signal processing problems, people manually select the typical signal samples to build the dictionary and let the algorithm run the training and update stage. Apart from these pure data-driven approaches, we propose a novel method which uses the features generated from the original signal in this paper. Considering that the statics feature generation methods, which usually use only one feature selection metric with fixed thresholds, may not always achieve the reliable performance on earthquake detection with various application scenarios~\citep{Langley-1994}, we take the strategy of generating adequate raw features from fixed length original signal with well-developed feature generation methods as well as the Python signal feature generation package, TsFresh~\citep{tsfresh}, and then followed by a combination of feature selection methods (including Relief-F~\citep{Rob-2003}, Gini Index~\citep{Gas-1972}, KL-Divergence~\citep{Kul-1951} and L1-Norm Regular Term~\citep{Ng-2004}) to pick out the features with the largest potential contribution. Another advantage brought by feature-based methods is that a bridge can be built between classical data-driven methods and traditional physical model analysis. Since all these features have clear physical meanings, analyzing the selected features could provide some hints for future physical research in earthquake detection.

In experiment and evaluation stage, we take the orthogonal matching pursuit~(OMP) and the simultaneous orthogonal matching pursuit~(SOMP) strategy to solve the dictionary learning problem and show that our results outperform the classical machine learning method --- support vector machine (SVM) \citep{hearst1998support}, as well as the extensively used earthquake detection method --- template matching (TM) \citep{gibbons2006detection}. To improve the detection accuracy, we slide one-fifth of the length of cutting window which is used to cut the original signal and produce features, which means most of the signal fragments (other than the fragments near the edge) will be classified for 5 times. A voting policy can then be applied. Only the signal fragments which are classified as positive for 3 or more times will be considered as a part of detected earthquake events. This voting strategy is also helpful in smoothing the detection results. All these machine learning methods are applied on the Labquake Dataset from Penn State University~\citep{wu2018deepdetect}, and the performance is evaluated by window-based accuracy and intersection over union (IoU) \citep{Far-2015}.

Our main contributions in this paper can be summarized as follows:
\begin{enumerate}
\item We develop a method of combining multiple feature selection metrics.
\item We introduce dictionary learning methods into earthquake detection task and compare its performance with other earthquake detection methods.
\item Based on dictionary learning classification, we further improve the detection accuracy with overlapping windows and voting techniques.
\end{enumerate}

\section{Related Works}
\label{Related Works}
Our work closely relates to the research in traditional earthquake detection, feature selection techniques, and dictionary learning. We first provide some background on each of those topics below.\\
\textbf{Earthquake Detection} \ During last decades, many seismological methods have been developed to detect and analyze earthquake events in 1-D time series data. Some of them build the wave function models to simulate the physical process of earthquake waves spreading in the lithosphere, and further study the patterns and properties \citep{Kwi-2016}. Some other detection methods were derived from signal processing and statistical methods. In seismology, Short Term Averaging/Long Term Averaging (STA/LTA) trigger is the most widely used detection method, because of its convenience in certain applications, such as weak-motion seismology \citep{trnkoczy1999topic}. It computes the ratio between the average absolute amplitude of a short-term window and a long-term window sliding on the seismic signal \citep{All-1982,Wit-1998}. Compared with STA/LTA, the autocorrelation method is able to achieve a better detection rate and is more suitable for detecting weak seismic signals \citep{Bro-2008}. The autocorrelation approach is known as a ``many to many'' detection technique. It searches for the objective waveforms via similarity without needing to provide
the desired waveform in advance, but is computationally inefficient. Compared with previous methods, template matching is developed to achieve a good balance between detection accuracy and computational efficiency, and a customized detection threshold which depends on the existing earthquake catalog and statistical information is employed to control the bias of this balance \citep{Gib-2006,She-2007}. Template matching computes the correlation coefficient between the sample waveform and the expert-selected templates through an ``one to many'' strategy. SVM is a well-developed machine learning method that was recently introduced in seismology to detect earthquakes~\citep{ruano2014seismic}. Given training samples, the classic SVM algorithm outputs an optimal hyperplane~\citep{hearst1998support} categorizing the samples in two classes, and this property makes SVM especially suitable for discriminating earthquake events from non-earthquake events.

\noindent \textbf{Feature Selection} \ Feature selection techniques are desired, because they help to reduce the interference from the irrelevant or redundant information in raw features, compress the dimension of input data, and maintain the accuracy of prediction models. At the early stage of machine learning, feature selection highly depends on domain knowledge \citep{Guyon-2002}. With contributions from information theory and machine learning techniques, various feature selection methods were developed and can be separated into two categories: filter methods and wrapper methods \citep{Kohavi-1997}. Filter methods basically rank all the features according to certain correlation criteria, and select the representative ones with manually-set threshold. Independent from the learning stage, these methods are generally adopted within the data preprocessing stage, so that the high efficiency and low computational complexity can be guaranteed \citep{Langley-1994}. Wrapper methods wrap around different combinations of features and evaluate their effectiveness through the performance of learning algorithm on validation set \citep{Guyon-2002}, which are different from the filter feature methods. Through evaluating within the learning and updating stage, the features selected by wrapper methods are specially optimized for certain type of learning algorithms. Both filter methods and wrapper methods are involved in the feature selection method proposed in this paper. Relief-F, Gini Index, Kullback–Leibler Divergence \citep{Kul-1951} and Subspace Detection \citep{harris2006subspace} , which are all examples of filter methods, select the potentially ``useful'' features according to the static criteria and thresholds. The L1-norm regularization in the object function of dictionary learning serves as a wrapper method. It finishes the selecting process training on the real data, and the features are selected in accordance with the performance of learning algorithm.
\\
\noindent \textbf{Dictionary Learning}
As a branch of sparse representation, dictionary learning methods aim at reconstructing the input signal with the sparsest representation, while minimizing the approximation error between the original signal and the signal reconstructed through dictionary atoms \citep{Karl-2010}. The dictionary is composed by the vector elements (also known as atoms) which are used to represent the input data. There are two typical groups of dictionary learning methods: greedy pursuit algorithms and convex relaxation algorithms. The goal of most greedy pursuit algorithms, such as the matching pursuit (MP) algorithm \citep{Mallat-1993} and the orthogonal matching pursuit (OMP) algorithm \citep{Tropp-2004}, is to find the locally optimal representation vectors in each iteration. On the other hand, convex relaxation methods transform the classic sparse representation problem into a strongly correlated convex problem. Many widely-used algorithms belong to this category, such as the basis pursuit denoising \citep{Chen-1999} and least absolute shrinkage and selection operator (LASSO) \citep{Tib-1996}. Some researchers have already applied dictionary learning methods for denoising and signal compression on seismic data \citep{zhu2015seismic}. To the best of our knowledge, incorporating dictionary learning with feature selection for earthquake detection tasks still remains as an unexplored area.

\section{Feature Generation and Selection}
\label{Feature Generation and Selection}

\subsection{Feature Generation}
During the process of generating features, we take the experiential features that have been used in time-series detection tasks into consideration, and we also take full advantage of the convenience brought by the general-purpose feature generation package in Python (TsFresh)~\citep{tsfresh} to produce as rich candidates as possible for the following selection step. In total, we generate 991-dimension raw features (a feature vector with 991 elements) for each input signal sample.\\
\textbf{Experiential Features: } According to previous works in earthquake detection and other tasks in 1-D time series data ~\citep{greene2008comparison,dzwinel2005nonlinear}, we pick 126-dimension features from time and frequency domain, such as maximum amplitude, number of peaks, and peak frequency. Some of these features may have potential in localizing or detecting the small anomalous earthquake events in time series data ~\citep{greene2008comparison}.\\
\textbf{Tsfresh Features: } Tsfresh is a python package which extracts abundant features to describe or cluster time series. This package generates 83 types of the features from frequency, power and entropy of the signal. The total dimension of the generated features is 865 (some of these features, such as continuous wavelet coefficients and energy of frequency bands, have multiple dimensions)~\citep{feature_list}.

\subsection{Feature Selection}

We combine the following feature selection methods together to obtain the key features: \\
\noindent \textbf{Relief-F: } Relief-F is an improved version of Relief algorithm. Relief algorithm measures the significance of a feature by its ability to distinguish neighboring instances. If feature distance between data points of same classes is large, it is less useful and gains a low weight. In contrast, if feature distance between data points of different classes is large, it is more useful and will gain a high weight. The final rank depends on the weight of each feature. Relief-F improves Relief through applying the k-nearest neighbors (KNN) method to each class \citep{Rob-2003}.
\\
\noindent \textbf{Gini Index: } Gini Index (GI, also known as Gini Coefficient of Inequality) is mainly used to evaluate the inequality within people’s wealth, population, etc., but it can also be applied in feature selection. Here we use GI to measure the ability of a feature to differentiate between target classes. GI is defined as the mean of absolute differences between all pairs of individual observations of the same feature
\begin{equation}
\displaystyle GI=\frac{\sum_{i=1}^{n}\sum_{j=1}^{n}\abs{x_{i}-x_{j}}}{2n^2\bar{x}},
\end{equation}
where $x_i$ and $x_j$ are the same feature observations of the different samples,  n is the number of values observed and $\bar{x}$ is the mean value. The minimum value of GI is 0 when all elements are equal and the theoretical maximum is 1, which is the ultimate inequality \citep{Gas-1972}. When GI is small, it means the feature basically provides similar information for different samples and is not helpful to distinguish the samples form various target classes. Therefore, features are ranked according to their GI, and the ones with high GI are usually selected since these features contain more useful information.
\\
\noindent \textbf{Kullback-Leibler Divergence: } The Kullback–Leibler Divergence (KL Divergence) \citep{Kul-1951} is also called Relative Entropy. It is used to measure how one probability distribution diverges from the second, expected probability distribution. Suppose $P(x)$ and $Q(x)$ are two discrete probability distributions of random variable vector x, the KL divergence between $P(x)$ and $Q(x)$ is
\begin{equation}
D_{KL}(P||Q)=-\sum_{i}P(x_i)\text{log}\frac{Q(x_i)}{P(x_i)},
\end{equation}
where $x_i$ is the i-th element in random variable vector x. This metric describes the reduction in entropy while replacing the feature which follows the distribution of $P(x)$ with another $Q(x)$-distributed feature. Thus, the feature which has larger KL divergence with other features are more likely to be discarded to avoid the redundant information.
\\
\noindent \textbf{Sparse Regression: } Sparse Logistic Regression(SLR) is an embedding feature selection algorithm utilizing L1-norm regularization. L1-norm can be considered as the optimal approximation of L0-norm. Using L1-norm as a regularization term encourages the sum of the absolute values of the parameters to be small. The number of training examples needed to achieve good performance grows logarithmically with the number of irrelevant features \citep{Ng-2004}. In SLR, L1-norm regularization is added to loss function so that the less useful attributes are given small weights. As a result, important features with high weights are picked out during the logistic regression.

\section{Dictionary Learning}
\label{Dictionary Learning}
In the field of dictionary learning, the input data can be represented in a sparse way through a linear combination of a group of basis vectors. These vectors (also known as the atoms in dictionary learning) compose the dictionary in the form of a 2D matrix, and they are usually used for signal representation \citep{mairal2009supervised}. The representation can be exact or be an approximation to the original signal. The dictionary learning methods usually build an over-complete dictionary matrix $ D \in R^{n \times K} $ which contains $ K $ column atoms with $ n $ elements respectively, and each column is 
initialized by an input sample. $ y \in R^n $ is the signal which needs to be represented as a sparse linear combination of different column atoms \citep{Michal-2006}. The representation of $ y $ can be written in the exact form $ y = Dx $, or in approximated form $ y \approx Dx $, which should satisfy the restriction $ \| y - Dx \|_p \leq \epsilon $ (where $\| \cdot \|_p $ represents the $ Lp $ norm, $\epsilon$ is the parameter used to control the residual error, and in this paper we manually set the value of $ p $ as 2). Each element in the vector $ x \in R^K $ can be considered as the coefficients of the input signal $ y $, so that the object of dictionary learning can be summarized as
\begin{equation}
\min_x \{ \|x\|_0 \}, \quad \text{s.t.}  \quad y = Dx,
\end{equation}
\noindent or 
\begin{equation}
\min_x \{ \|x\|_0 \}, \quad \text{s.t.}  \quad \| y - Dx \|_2 \leq \epsilon.
\end{equation}

In recent years, a number of methods have been developed to solve the dictionary-learning problems. One of the most direct methods is the orthogonal matching pursuit~(OMP) algorithms~\citep{Tropp-2004}. One of the crucial part of OMP algorithm is introducing the $ L_2 $ norm regularizer to normalize each atom (each column) in the dictionary. Let $D_i$ represents the i-th column in dictionary matrix $D$, so that $\|D_i\|_2 = 1$ for $i = 1,2, \ldots,K$. $ D(S) $ is a sub-matrix of $ D $, which consists of the $ith$ columns of $ D $ with $ i \in S $ and $ S \subset \{ 1,2, \ldots, K \}$. In each iteration of OMP, the algorithm is able to pick out the atom which has the most important contribution in reconstruction from all the candidate vectors and assign it to our dictionary. The detailed description of OMP algorithm can be stated as follows
\begin{enumerate}
\item Use the input vector $ y $ to initialize the residual $ r_0 $, and initialize the selected variables set as $ D(c_0) = \emptyset $, where $ c_i $ is a set of atom indexes. Set the iteration counter variable $ i = 1 $.
\item Find the atom $ D_t $ to satisfy the maximization function
\begin{equation} 
\max_t | D_t^T r_{i-1} |,
\end{equation}
and add the selected atom $ D_t $ to the set of selected atoms. Update $ c_i = c_{i-1} \cap \{ t_i \} $.
\item Update $ P_i=D(C_i)(D(c_i)^T D(c_i))^{-1}D(c_i)^T $, where $P_i$ denotes the projection onto the linear space spanned by the atom vectors in $D(c_i)$. Update the residual $r_i = y - \hat{y} = (I - P_i)y $, which is defined as the difference between input vector $y$ and the vector $\hat{y} = P_i y$ after projection.
\item If the stopping iteration time has been achieved, stop the algorithm and return the residual $ r_i $ and selected set $ D(c_i) $. Otherwise, let $ i = i + 1 $ and return to Step 2.
\end{enumerate}
\noindent The simultaneous orthogonal matching pursuit (SOMP) algorithm basically shares the structure with OMP algorithm. The only difference is using a set of input samples $ y \in R^{m \times K} $ instead of the single input sample $ y \in R^K $. Inspired by the dictionary learning model built by Determe et al. \citep{det-2015}, we use 6 neighborhood samples (3 samples on the left and 3 samples on the right) and the original sample to build the input matrix $y$. Due to the fact that the fixed length cutting windows may not cut out the entire earthquake signal, we include some neighboring information with SOMP algorithm to further improve the detection performance of our model.

While applying dictionary learning to binary classification problem such as the earthquake detection task, we need to build one dictionary containing feature vectors generated from positive samples (which contain over 50\% time samples labeled as earthquake events) and another dictionary consisting of features from negative samples (which contain over 50\% time samples labeled as non-earthquake events). After training, we will compare the residuals of testing samples on positive and negative dictionary, and classify the test samples to the class with smaller residuals.

\section{Experiments}
\label{Experiments}

\subsection{Dataset}

\begin{figure}
\centering
\noindent\makebox {
\begin{tabular}{c}
\includegraphics[width=0.8\linewidth]{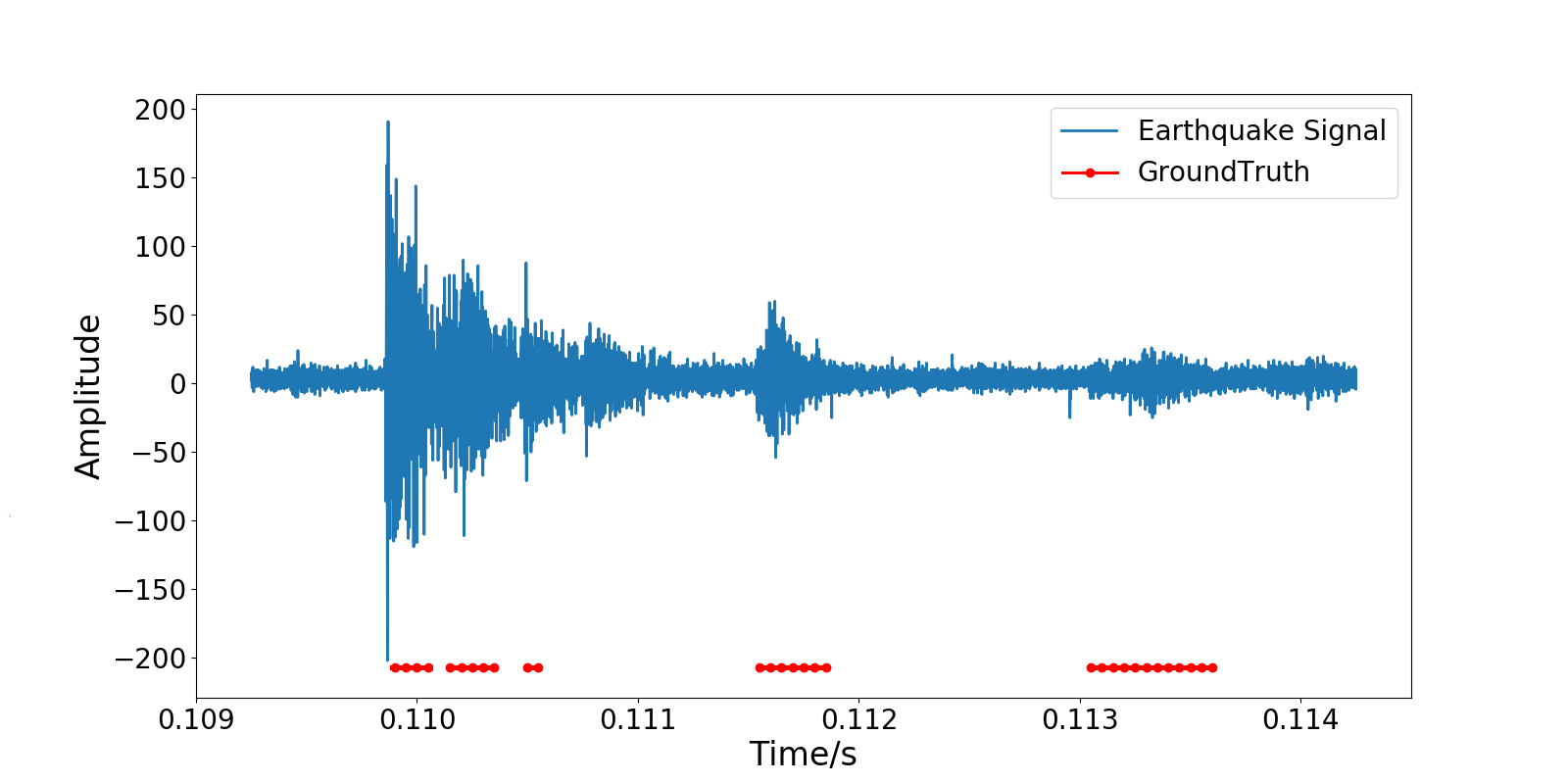} \\
(a)\\
\includegraphics[width=0.5\linewidth]{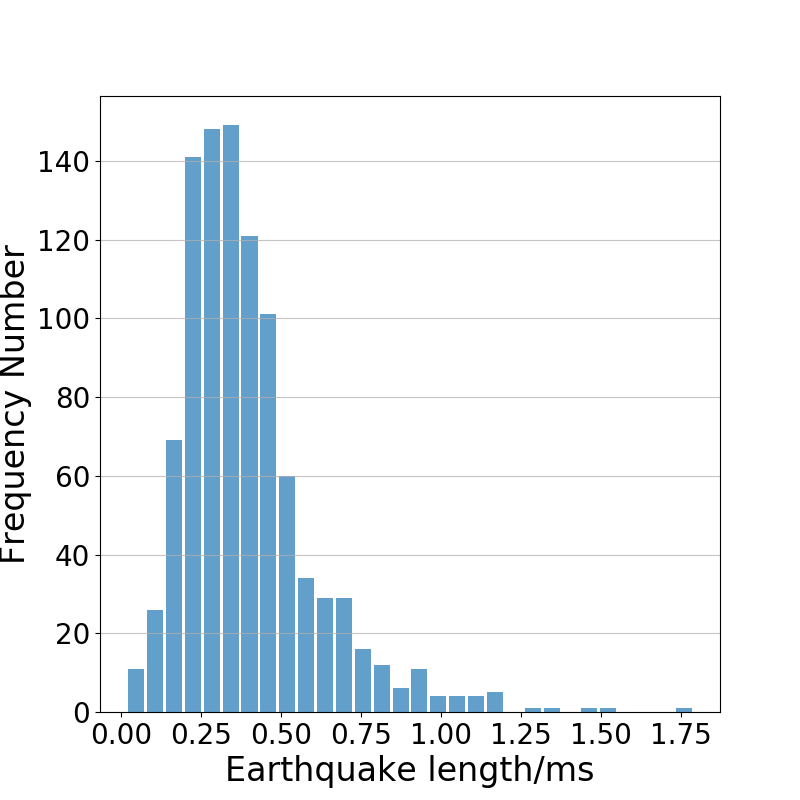}\\
(b)\\
\end{tabular}
}
\caption{ (a) A fragment of labquake signal, where the red points approximately demonstrate the start and end points of the earthquakes manually labeled by experts. There are 200 time samples between two continuous red points; (b) A histogram shows the distribution of earthquake event length. }
\label{Fig:Earthquake_signal}
\end{figure}

\begin{table}[!ht]
\centering
\begin{tabular}{|c|c|c|}
\hline
\textbf{Statistical Information} & \textbf{time samples} & \textbf{Time(s)}\\
\hline
\hline
Maximum Length of Earthquake Events & 7,151  & $1.79 \times 10^{-3}$\\
\hline
Minimum Length of Earthquake Events & 74  & $1.85 \times 10^{-5}$ \\
\hline
Average Length of Earthquake Events & 1,275  & $3.94 \times 10^{-4}$\\
\hline
Median Length of Earthquake Events & 1,397  & $3.49 \times 10^{-4}$\\
\hline
75\% Percentile of Earthquake Events & 1,893  & $4.73 \times 10^{-4}$\\
\hline
80\% Percentile of Earthquake Events & 2,019  & $5.05 \times 10^{-4}$\\
\hline
90\% Percentile of Earthquake Events & 2,643  & $6.61 \times 10^{-4}$\\
\hline
Total Length of Earthquake Events & $1.27 \times 10^6$  & 0.32 \\
\hline
Total Length of Non-earthquake Events & $2.08 \times 10^6$  & 0.52  \\
\hline 
\end{tabular}
\caption{Some statistical information about the length of the earthquake events in the labquake dataset.}
\label{Tab:Event_length}
\end{table}

Our labquake dataset is provided by the Rock and Sediment Mechanics Laboratory at Penn State University. The labquake data is a time-amplitude representation generated by a machine to mimic real seismic signals. Overall, there are 3,357,566 time samples in our dataset. Two broadband piezoelectric transducers are used to record the earthquake acoustical data, and the labquake data are sampled continuously at 4 MHz with a 14-bit Verasonics data acquisition system \citep{labquake}. 1,000 seismic events are manually picked, and we use 800 events for training, 100 events for validating, and 100 events for testing. Since the noise within the data generation process has been controlled at a relatively low level, no additional denoising filter is applied in our experiments.

A fragment of the labquake event is shown in Fig.~\ref{Fig:Earthquake_signal}(a), and we gather the statistical information of all the event length in Table~\ref{Tab:Event_length}. The histogram of event length shows the length distribution in Fig.~\ref{Fig:Earthquake_signal}(b). Since most of the length of earthquake events are smaller than 0.5 ms (2,000 time samples), we choose 2,000 time samples as the length of the sliding window for our detection algorithm. As shown in Table~\ref{Tab:Event_length}, the longest event lasts more than 7,000 time samples, while the shortest one only spans less than 100 time samples. The maximum absolute amplitude of the largest earthquake event achieves 756, while the absolute peak amplitude of the smallest event is 14. The average of the maximum absolute amplitude of all the earthquake is 107.

\subsection{Generated and Selected Features}

At this stage, we cut the original signal into 2,000-time-sample-long sections without overlap. If over 50\% time samples are labeled as earthquake events, this section can be considered as a positive sample, otherwise, it will be considered as a negative sample. We generate the 991-dimension features in total with the methods described in Section 3. Three feature selection methods (Relief-f, Gini Index and KL-divergence) are applied to evaluate the potential contribution of each feature to our detection model. Under each of these selection methods, a score can be assigned to these features to evaluate their potential importance. We manually set the selection metric as discarding the features which have the worst 25\% performance under each feature selection metric, and 507-dimension features are left. The detailed information and comparison have been summarized in Table~\ref{Tab:features}.

From Table~\ref{Tab:features}, the experiential features have a higher selection rate compared to the features from Python signal package TsFresh, which demonstrates that features based on human experience may have some advantages in 1D time series detection task. Among the features in four different domains (time, frequency, energy and others), features generated from frequency domain contribute the most; they take a significant proportion in all the features and also have a higher tendency to be selected by our feature selection methods.

\begin{figure}[!t]
\centering
\noindent\makebox {
\begin{tabular}{c}
\includegraphics[width=0.6\linewidth]{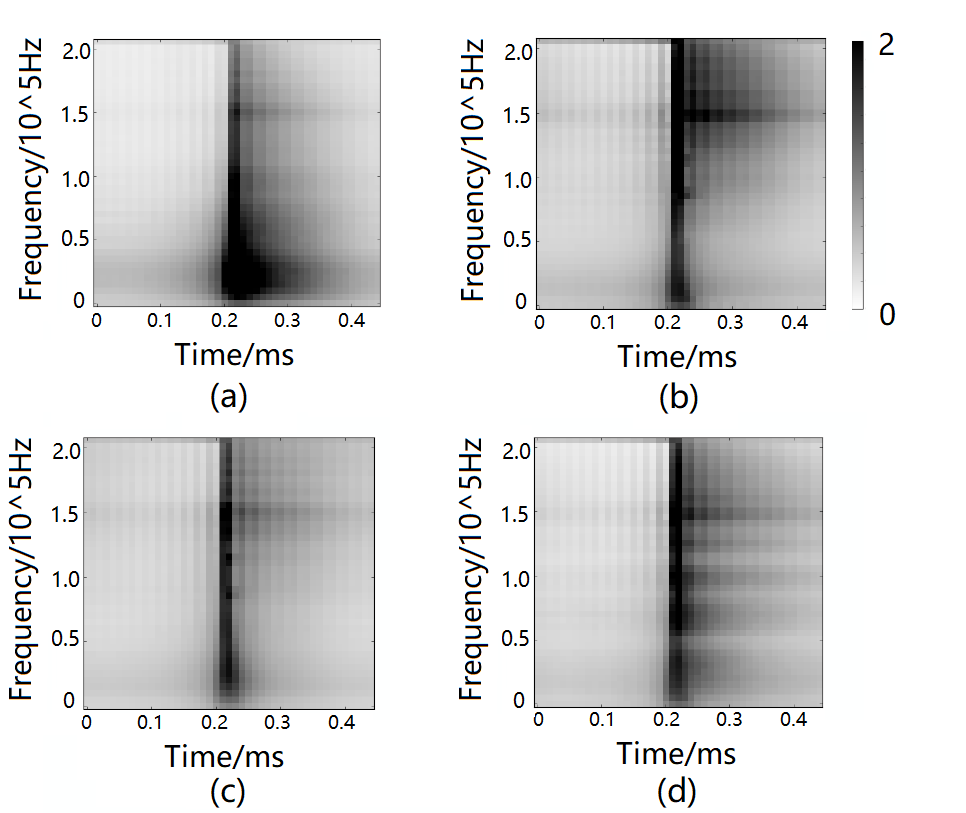} \\
\end{tabular}
}
\caption{(a)(b) The logarithm-based time-frequency graph of two sections of signal which contain earthquake events; (c)(d) The logarithm-based time-frequency graph of two sections of signal which contain non-earthquake events.
}
\label{fig_TFPlot}
\vspace{0.5cm}
\end{figure}

To further exploit this point, we apply Logarithm-Based Time-Frequency analysis \citep{Brian-1990} to both positive and negative samples, and plot the results in Fig.~\ref{fig_TFPlot}. In particular, Figs.~\ref{fig_TFPlot}(a) and (b) show the time-frequency graphs of a big earthquake event and a small earthquake event, respectively (small events usually indicate the earthquake events that have smaller amplitude and last for shorter time, while big events have the opposite properties.). Figures~\ref{fig_TFPlot}(c) and (d) demonstrate two examples from negative samples which do not contain earthquake events. From these time-frequency graphs, the frequency components of earthquake samples tend to concentrate in certain frequency bands since the  pattern of the labquake dataset is not complicated, and it does not contain much noise components. In contrast, the time-frequency graphs of negative samples do not seem to contain such property. 

\begin{figure}[!ht]
\centering
\includegraphics[width=5in]{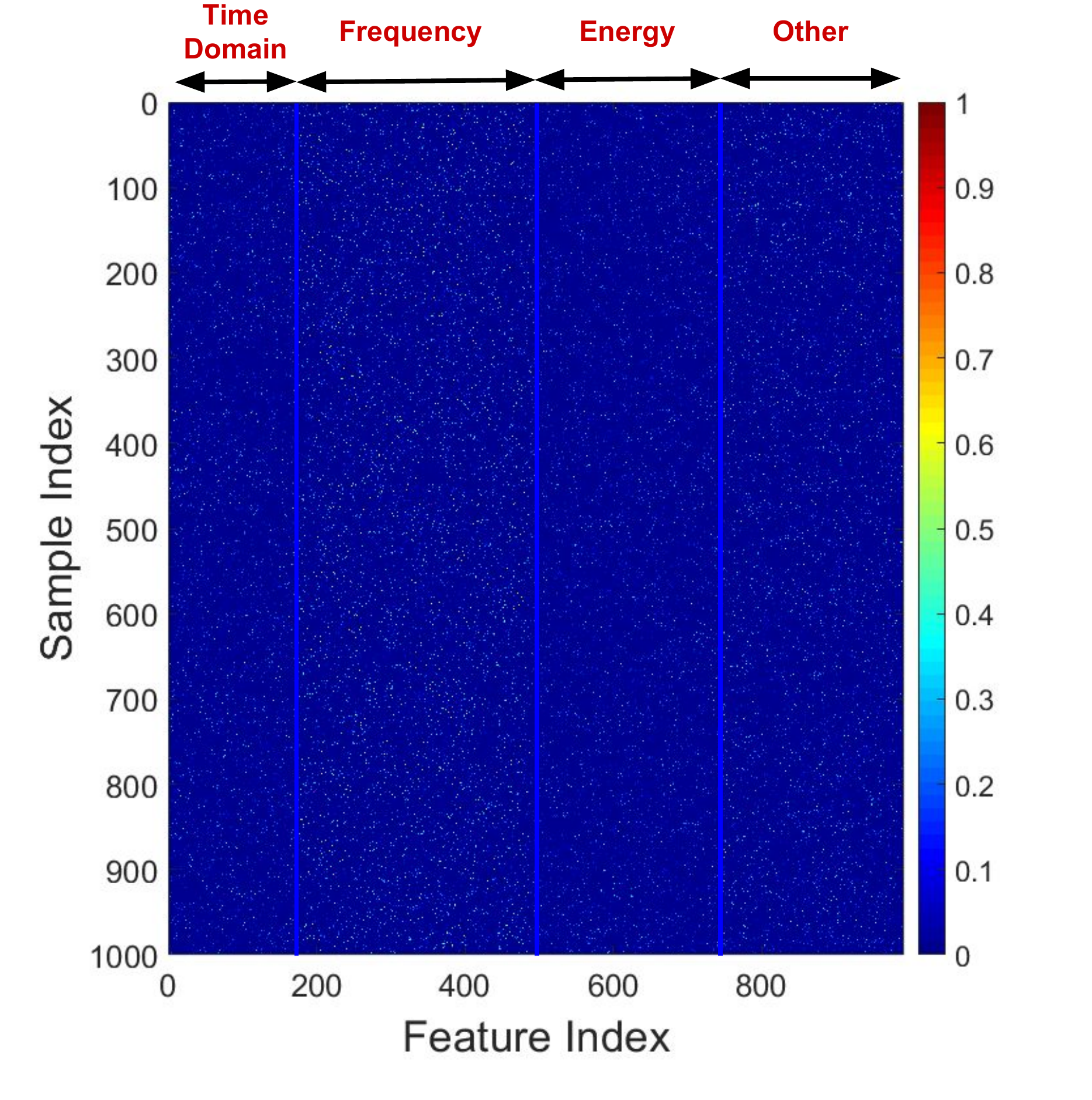}
\caption{The illustration of wrapper selection results of randomly picked 1,000 signal samples from our dataset, where feature index in Columns~1-172 represent time domain features. Columns~173-487 are frequency domain features. Columns~488-744 are features related with energy. Columns~745-991 are features which do not belong to previous types. Each row represents the weight vector generated by wrapper selection method, and the elements in a weight vector correspond to the importance of 991-dimension features derived from each 2,000-sample signal sample. The colors represent the relative importance of different features, with the light-colored scattered elements having the highest importance.
}
\label{fig_selection}
\vspace{0.5cm}
\end{figure}

Then, we illustrate the normalized weight vectors (row vectors) of 1,000 randomly selected signal samples generated from wrapper selection method in Fig.~\ref{fig_selection}. Each column represents the weight elements of the same feature in different samples. Features in Columns~1-172 are derived from time domain. Columns~173-487 represent frequency domain features. Columns~488-744 are energy-related features. Columns~745-991 represent the features generated from other methods. Due to the existence of L1-norm regularization from SLR, the corresponding elements in weight vector of the unimportant features are set to zeros. The scattered elements with light colors represent high weight values, and they can be considered as the relatively important features picked by wrapper method. It can be seen that the frequency domain features tend to have a higher selection rate. This result indicates these features play more important roles than others in our detection model, and it is consistent with the selection result of filter methods. These frequency-related properties may have some potential values in further physical model analysis\citep{Boj-2013}.

\begin{table}[!ht]
\small
\centering
\begin{tabular}{|c|c|c|c|c|c||c|c|}
\hline
  & Time Domain & Frequency  & Energy & Other & Total Selected & Total Generated & Selection Rate\\
\hline
Experiential & 18 & 32 & 16 & 7 & 73 & 126 & 57.9\%\\
\hline
TsFresh & 65 & 156 & 85 & 118 &  434 & 865 & 50.1\%\\
\hline
Total Selected & 83 & 188 & 101 & 125 & 507 & 991 & 51.2\%\\
\hline
\hline
Total Generated & 172 & 315 & 257 & 247 & & & \\
\hline
Selection Rate & 48.3\% & 54.5\% & 39.3\% & 50.6\% & & & \\
\hline
\end{tabular}
\caption{Table of Feature Selection Results}
\label{Tab:features}
\end{table}

\subsection{Detection Results}

We apply two different quantitative metrics to evaluate the performance of our detection model: the window-based accuracy (abbreviated as ``accuracy'' in the following part) and intersection over union (IoU, also known as Jaccard similarity) \citep{Far-2015,yoon2015earthquake}. The window-based accuracy is the accuracy of detecting whether the continuous signals segmented by fixed-length windows contain earthquake events. This metric can be considered as the evaluation of overall detection results. In contrast, intersection over union (IoU), defined as
\begin{equation}
IoU = \frac{Detections \cap Ground \ Truth}{Detections \cup Ground \ Truth}.
\label{eq:iou}
\end{equation}
Intersection over union (IoU) especially focuses on the detection performance on positive samples. Since the length of negative samples dramatically exceed the positive samples in our dataset (as shown in Table~\ref{Tab:Event_length}, the ratio between the total number of time samples of negative samples and positive samples is about 5:3), the overall evaluation is more inclined to be affected by the prediction accuracy on negative samples. Considering the fact that researchers are usually more interested in the detection results on positive samples which contain earthquake events, IoU is adopted as the evaluation metric to balance the importance of positive and negative samples in learning process.

\begin{table}[!ht]
\centering
\begin{tabular}{|c|c|c|}
\hline
\textbf{Detection Methods}  & \textbf{Accuracy} & \textbf{IoU} \\
\hline
\hline
Template Matching & 48.7\% & 0.12 \\
\hline
Raw Signal + SVM & 49.4\%  & 0.14 \\
\hline
Raw Features + SVM & 54.8\% & 0.19 \\
\hline
Raw Signal + OMP & 52.7\% & 0.17 \\
\hline
\hline
Relief-F + OMP & 63.5\% & 0.23 \\
\hline
Gini Index + OMP & 69.4\% & 0.27 \\
\hline
KL divergence + OMP & 71.8\% & 0.30 \\
\hline
Combination Selection + OMP & 73.7\% & 0.33 \\
\hline
Combination Selection + SOMP & 78.3\% &  0.38\\
\hline
Combination Selection + SOMP + Voting Strategy & 80.1\% & 0.42 \\
\hline
\end{tabular}
\caption{An illustration of the detection results of with the different combination of feature selection methods and dictionary learning strategies. All the detection results are assessed under the criteria of the average Accuracy and IoU after 5-fold cross-validation.}
\label{Tab:detection}
\end{table}

\begin{figure}[!t]
\centering
\noindent\makebox {
\begin{tabular}{c}
\includegraphics[width=0.85\linewidth]{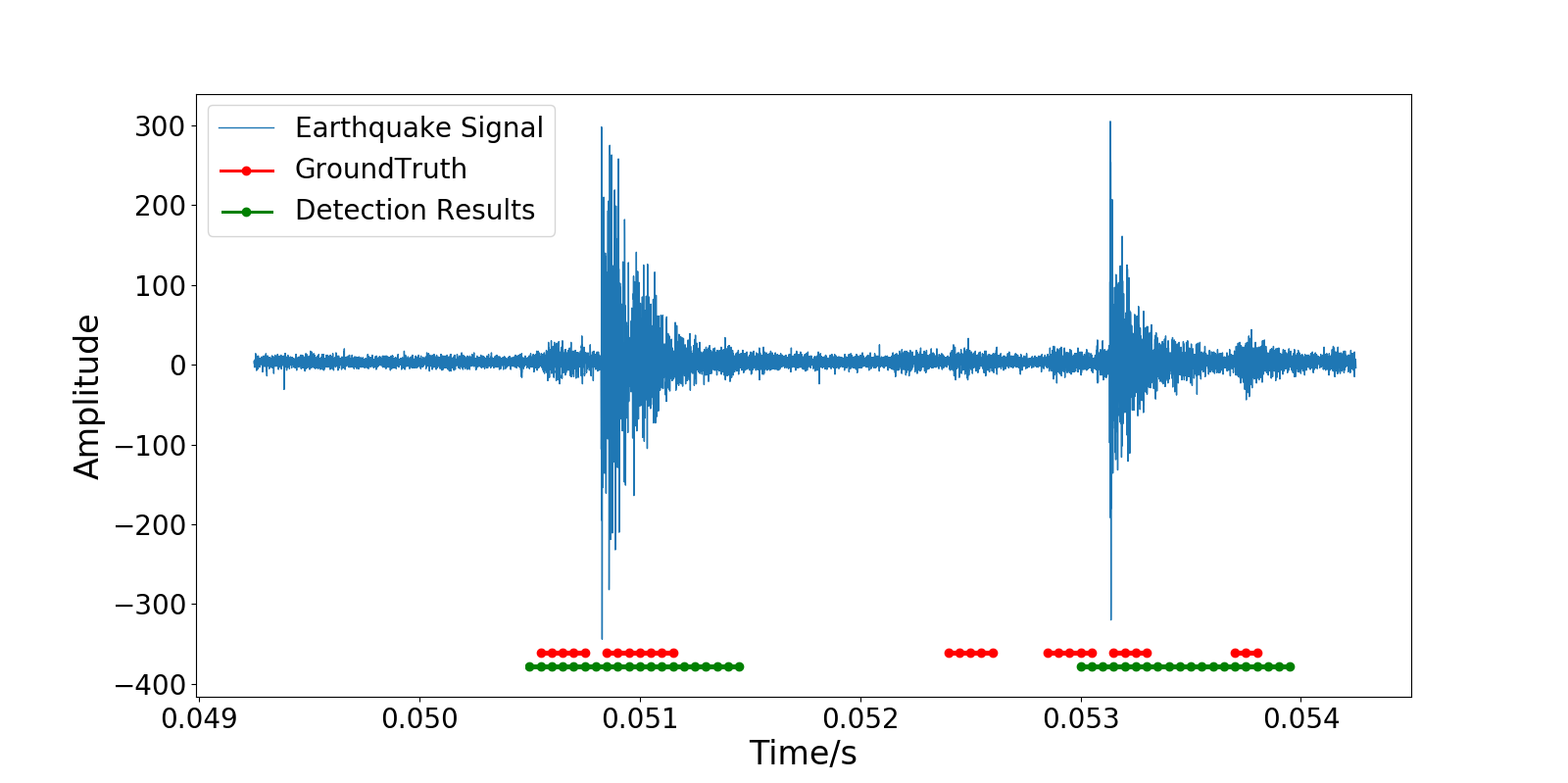} \\
(a)\\
\includegraphics[width=0.85\linewidth]{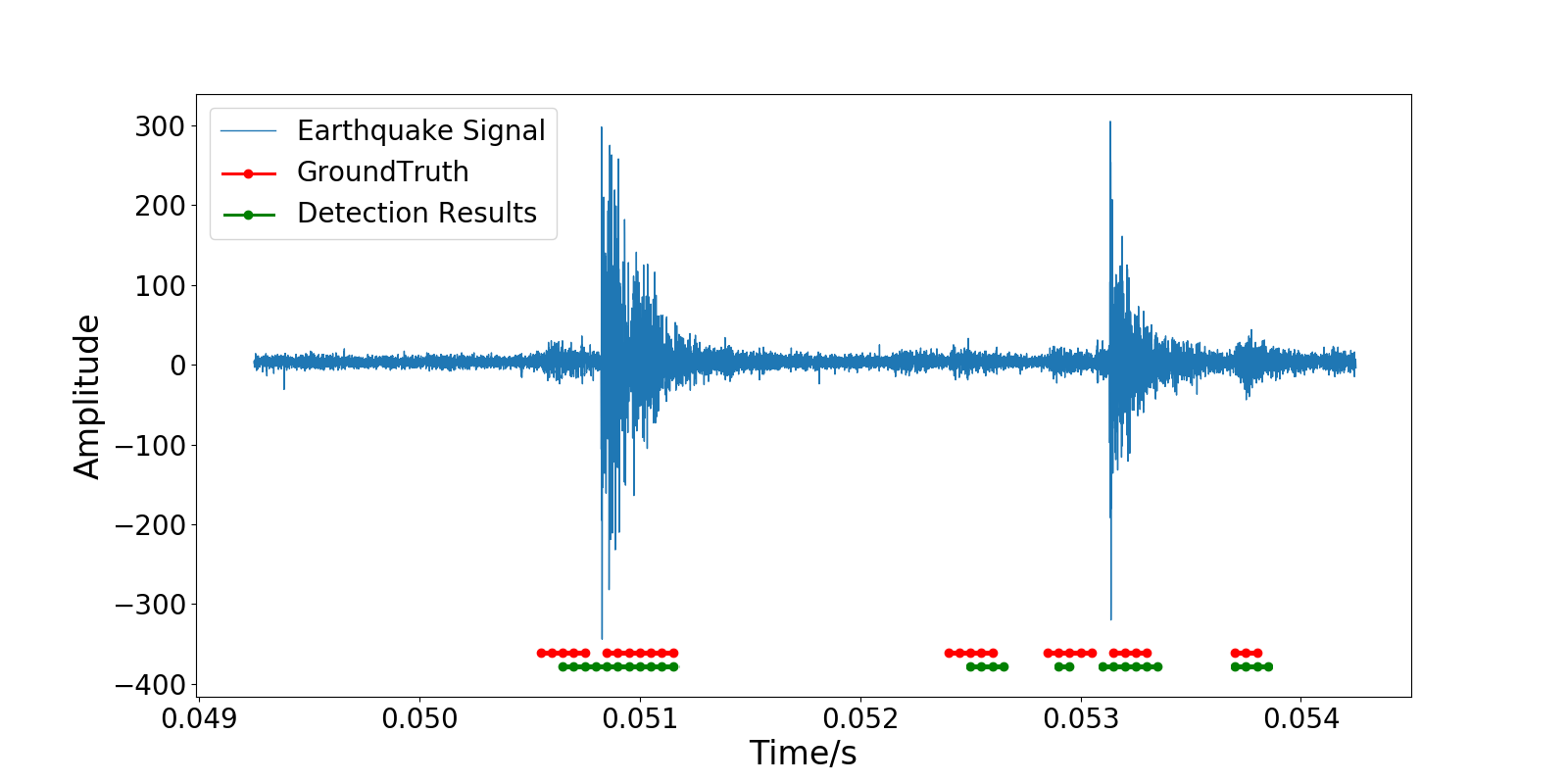}\\
(b)\\
\end{tabular}
}
\caption{ (a) The detection result without overlapping windows, where the resolution of detection is 2,000 time samples ($5 \times 10^{-4}$ second). (b) The detection result with four fifth overlapping windows, where the resolution of detection is 400 time samples ($1 \times 10^{-4}$ second). In both (a) and (b), the blue curve denotes original signal, red points represent ground truth and green points represent prediction results. There are 200 time samples between two neighboring points.}
\label{fig_detection_events}
\end{figure}

In Table~\ref{Tab:detection}, we compare the performance of different combinations of feature selection methods and dictionary learning strategies with traditional template matching and SVM. The template matching methods we used follow the framework proposed by~\citet{gibbons2006detection} and the parameters are chosen similar to the work completed by~\citet{wu2018deepdetect}. To reduce the effect of randomness, the technique of k-fold cross validation is used in the experimental stage. This technique first separates the dataset into k evenly sized subsets, and then iteratively picks one subset for validation and testing and the other four for training~\citep{kohavi1995study}. In our experiment, we manually set k to 5, and the average detection accuracy and IoU over $k=5$ sbusets are used to evaluate the performance to increase the robustness. As shown in Table~\ref{Tab:detection}, Support Vector Machine (SVM) may not be suitable for the classification task with high feature dimensions. SVM tends to overfit during the training process, which means the detection model achieves good detection results on training samples but poor results on testing data. Template matching yields  detection accuracy comparable to SVM. Considering that our detection task is a binary classification problem, the detection results with accuracy lower than 60\% (the first four methods in Table 4) is not significantly better than a random guess, and we consider these as failed detections. Generally, our dictionary learning methods could achieve more reliable detection results than template matching and SVM methods. Among all the dictionary learning methods, directly fitting the original signal into dictionary performs the worst, since the amplitude of our earthquake signal data could vary within wide limits, and the amplitude of some small events close to big events would approach to zero after normalization. The models based on the combination of feature selection methods significantly outperform the models based on single feature selection method and raw features. This demonstrates that the combination of multiple feature selection method can improve the robustness and accuracy of detection. The comparison between OMP and SOMP shows that the accuracy of these two methods are close, while SOMP yields better IoU value. This proves that taking the neighboring information into consideration plays a vital role in achieving a better detection result for earthquake events.

To further improve our detection results and achieve higher detection resolution, we apply the voting strategy. Instead of cutting the original signal into 2,000-time-sample-long windows without overlap, we slide the cutting window one-fifth of its length (400 time samples) every time, which means each time window has a $80\%$ (1,600 time samples) overlap with its previous one. We apply our classification models to these overlapping samples. Except for the beginning and the ending parts, most of the 400-time-sample-long signal fragments will be classified 5 times. We set the classification criteria that only the fragments classified as earthquake 3 or more times can be considered as positive. In this case, the detection resolution is successfully increased from 2,000 time samples to 400 time samples, and this model achieves the best accuracy. From Fig.~\ref{fig_detection_events}, the comparison between the detection results before and after applying the voting strategy also shows that this strategy is helpful in improving the accuracy. However, both Fig.~\ref{fig_detection_events}(a) and Fig.~\ref{fig_detection_events}(b) demonstrate our models still yield the low accuracy on detecting the small earthquake events close to large events. We suspect that this is because the features of these small events is closer to the background non-earthquake events while comparing with the features from the earthquake events with large amplitude. A detailed and comprehensive test of this hypothesis is necessary, and we leave this for future work.

\begin{table}[!t]
\centering
\begin{tabular}{|c|p{8cm}|}
\hline
\textbf{Experiential Features} & \textbf{Description}\\
\hline
\hline
Minimum Value  & The minimum value of the input signal.  \\
\hline
Maximum Value  & The Maximum value of the input signal.  \\
\hline
Root Mean Square Amplitude &  the square root of the average of the squared values of the in signal. \\
\hline
Zero Crossing &  The total number of intercept points with the time axis.  \\
\hline
Activity (Variance) &   The variance of the input signal.\\
\hline
Mobility &  The square root of the ratio of the variance of the first derivative of the signal to the activity of the original signal.  \\
\hline
Peak Frequency &  The frequency of maximum power in the frequency spectrum. \\ 
\hline
Spectral Edge Frequency (SEF) & The frequency below which 50\% percent of the total power of the input signal are located.  \\
\hline
 Total Spectral Power  & The sum of squared amplitude of the input signal. \\
\hline
Spectral Entropy & The complexity of the frequency spectrum of the input signal~\citep{misra2004spectral}  \\
\hline 
\end{tabular}
\caption{The demonstration and brief description of the most important (the most frequently-selected) features picked by our feature selection methods from the experiential features.  }
\label{Tab:exp_selected}
\end{table}

\begin{table}[!ht]
\centering
\begin{tabular}{|c|p{8cm}|}
\hline
\textbf{Tsfresh Features} & \textbf{Description}\\
\hline
\hline
 The Absolute Value of Consecutive Changes &  Sum over the absolute value of consecutive changes in the input signal. \\
\hline
Number of Peaks & Calculate the number of peaks of at least support n (parameter) in the signal. \\
\hline
 Count Above Mean & Count the number of values in the signal that are higher than the mean.  \\
\hline
 First Location of Maximum & The first location of the maximum value of the input signal.\\
\hline
 Skewness &  Calculate the skewness with the adjusted Fisher-Pearson standardized moment coefficient G1.~\citep{hillis1992signal} \\
\hline
 Fast Fourier Transform (FFT) Aggregated & Generate the spectral centroid (mean), variance, skew, and kurtosis of the absolute Fourier Transform spectrum~\citep{hsia2008bayesian}.  \\
\hline
FFT Coefficients & Calculate the spectrum of the input signal after Fast Fourier Transform.  \\
\hline
Absolute Energy & Sum the absolute values of the amplitude in input signal. \\
\hline
Continuous Wavelet Transform(CWT) Coefficients  & Calculates a Continuous wavelet transform for the Ricker wavelet~\citep{kallweit1982limits}. \\
\hline
 Number of CWT Peaks &  This feature calculator searches for different peaks in input signal~\citep{du2006improved}. \\ 
\hline
 Shannon Entropy &  Defined as the classic entropy in Information Theory~\citep{dincer2001energy}.  \\
\hline
 Approximate Entropy &   A coeffieient used to quantify the amount of regularity and the unpredictability of fluctuations over time-series data~\citep{pincus1991approximate}. \\
 \hline
  Binned Entropy    & The first bin value into the equidistant bins of the input signal.  \\
\hline 
\end{tabular}
\caption{The demonstration and brief description of the most important (the most frequently-selected) features picked by our feature selection methods from the Tsfresh features.  }
\label{Tab:tes_selected}
\end{table}

\section{Discussion}

In this paper, we develop a novel method for earthquake events detection based on feature selection and dictionary learning. In particular, we develop a few different computational techniques to improve the detection accuracy including a combination of different features, accounting neighboring information, and a voting strategy. We evaluate the effectiveness of our model through different metrics and compare its performance with other machine-learning-based detection methods as well as the traditional earthquake detection method. Results show that our detection method yields higher accuracy.

After the 5-fold cross validation experiments, our feature selection methods have been conducted 5 times. According to the voting strategy, the features which are most frequently-selected (selected 5 times) by our feature selection methods can be considered as the ``important'' ones. Tables~\ref{Tab:exp_selected} and \ref{Tab:tes_selected} provide a brief description of the most important (the most frequently-selected) features with clear physical meaning picked by our feature selection methods from the experiential features and Tsfresh features. Some of the features may have multiple dimensions, such as ``Fast Fourier Transform Aggregated'' and ``Continuous Wavelet Transform Coefficients''.  

One of the main limitations of our feature selection and dictionary learning incorporated model is that the voting strategy detection method is highly reliant on the size of overlapping signals. The accuracy is calculated on the binary classification with fixed window-length signal, which is usually not matched with the ground truth label and restricted the detection resolution. Even when the window-based (400 time samples) detection accuracy achieves 100\%, the IoU still remains as 0.63. To achieve higher detection resolution, a smaller sliding stride is needed. However, this may tremendously increase the computation complexity. More efficient feature selection and dictionary learning methods will become necessary. Currently, our experiments are conducted on the simulated and noise-free labquake data. To extend the application of our detection method to real earthquake detection, more experiments on the data with lower sampling rates, higher noise level and more non-earthquake events with complicated pattern will be tested in the future.

\section{Data and Resources}

We use time series data acquired at the Rock and Sediment Mechanics Laboratory of Penn State University. The dataset is a time-amplitude representation generated by a double-direct shearing apparatus to mimic real Earthquake. All the data sets are available for download  at \url{https://sites.psu.edu/chasbolton/} .

\section{Acknowledgment}
We thank two anonymous reviewers and the editor for providing valuable suggestion to improve the original manuscript. We thank for the Rock and Sediment Mechanics Laboratory at Penn State University for providing us the labquake dataset. This work was co-funded by the Center for Space and Earth Science at Los Alamos National Laboratory.

\bibliographystyle{IEEEtranN}
\bibliography{bibliography}

\end{document}